\documentstyle[prl,aps,preprint,epsfig,epsf,floats]{revtex}

\begin{document}           
\newcommand{\be}{\begin{equation}}
\newcommand{\ee}{\end{equation}}
\newcommand{\bq}{\begin{eqnarray}}
\newcommand{\eq}{\end{eqnarray}}
\newcommand{\Sc}{Schr\"odinger\,\,}
\newcommand{\Sp}{\,\,\,\,}
\newcommand{\no}{\nonumber\\}
\newcommand{\tr}{\text{tr}}
\newcommand{\p}{\partial}
\newcommand{\la}{\lambda}
\newcommand{\La}{\Lambda}
\newcommand{\G}{{\cal G}}
\newcommand{\D}{{\cal D}}
\newcommand{\E}{{\cal E}}
\newcommand{\W}{{\bf W}}
\newcommand{\de}{\delta}
\newcommand{\al}{\alpha}
\newcommand{\bi}{\beta}
\newcommand{\ep}{\varepsilon}
\newcommand{\ga}{\gamma}
\newcommand{\epp}{\epsilon}
\newcommand{\vep}{\varepsilon}
\newcommand{\th}{\theta}
\newcommand{\om}{\omega}
\newcommand{\si}{\sigma}
\newcommand{\J}{{\cal J}}
\newcommand{\pr}{\prime}
\newcommand{\ka}{\kappa}
\newcommand{\TH}{\mbox{\boldmath${\theta}$}}
\newcommand{\DE}{\mbox{\boldmath${\delta}$}}
\newcommand{\lan}{\langle}
\newcommand{\ran}{\rangle}
\newcommand{\Hol}{\text{Hol}}
\newcommand{\cp}{{\bf CP}}
\newcommand{\spp}{\,\,\,\,\,\,\,\,\,\,\,\,\,\,}

\bibliographystyle{unsrt}
 
\setcounter{page}{0}
\def\footnoterule{\kern-3pt \hrule width\hsize \kern3pt}
\tighten
\title{
Optical Holonomic Quantum Computer 
}
\author{Jiannis Pachos\footnote{Electronic address: pachos@isiosf.isi.it} and Spiros Chountasis\footnote{Electronic address: spiros@isiosf.isi.it}\\
{~}
}

\address{
Institute for Scientific Interchange Foundation, \\
Villa Gualino, Viale Settimio Severo 65, I-10133 Torino, Italy\\
{~}
}

\date{October 1999}

\maketitle

\thispagestyle{empty}

\begin{abstract}
In this paper the idea of holonomic quantum computation is realized within quantum optics. In a non-linear Kerr medium the degenerate states of laser beams are interpreted as qubits. Displacing devices, squeezing devices and interferometers provide the classical control parameter space where the adiabatic loops are performed. This results into logical gates acting on the states of the combined degenerate subspaces of the lasers, producing any one qubit rotations and interactions between any two qubits. Issues such as universality, complexity and scalability are addressed and several steps are taken towards the physical implementation of this model.
\end{abstract}



\section{Introduction}

Holonomic transformations have been recently proposed \cite{ZARA} and more extensively studied \cite{PAZARA} as logical gates for quantum computation \cite{QC}. The idea formal as it may apear at a first glance is not confined to a purely theoretical shere, but has the challenging possibility of experimental implementation. Towards this purpose we employ existing devices of quantum optics, such as displacing and squeezing devices and interferometers acting on laser beams in a non-linear medium. A different setting of optical quantum computer has been reported in \cite{Chuang}. The attempt to apply the abstract idea of holonomic quantum computation (HQC) to a physical system has lead us to deal with and clarify some theoretical problems of HQC such as universality, complexity (tensor product structure of qubits) and scalability. On the other hand, the experimental setup of the scheme proposed may prove to be a possible even though challenging task for the experimenters.

The basic idea of HQC is related with the geometrical phases \cite{SHWI} generated by the 
isospectral transformations of an $n$-fold
degenerate Hamiltonian, $H_0$, so its presentation can be given in a geometrical form \cite{ZARA}.
Initially, quantum information is encoded in the $n$ dimensional degenerate eigenspace $\cal C$ of $H_0$, 
with eigenvalue $ E_0$. The operator
$H_0$ is considered to belong to the family ${\cal F}=\{H_\sigma={\cal U}(\sigma)\,H_0\,
{\cal U}^\dagger(\sigma);\sigma\in{\cal M}\}$ of Hamiltonians unitarily (${\cal U}^\dagger (\sigma)={\cal U}^{-1}(\sigma)$) equivalent and therefore isospectral with $H_0$, where $H_0=H_{\sigma_0}$ for some $\sigma_0 \in {\cal M}$.
as $\sigma$ ranges over the control manifold $\cal M$ no energy level crossing occurs.
The $\sigma$'s represent the classical ``control'' parameters that one uses in order
to manipulate the encoded states $|\psi\rangle\in{\cal C}$.
Let $C$ be a {\em loop} in the control manifold $\cal M$.
When the loop $C$ is slowly gonne through, then no transition among different energy levels occurs
and the evolution is adiabatic, i.e. ${\cal F}$ is faithfully realized by the experimental setup. 
If $|\psi\rangle_{in}\in{\cal C}$ is an initial state in the degenerate 
eigenspace, at the end of the loop it becomes
$ |\psi\rangle _{out}=e^{i\,E_0\,T}\, \Gamma_{A}(C) \,|\psi\rangle_{in}$. 
The first factor is just an overall dynamical phase which in the following 
will be omitted by a redefinition of the energy levels, taking $E_0=0$. 
The second contribution is the holonomy $\Gamma_{A}(C)\in U(n)$, 
and is a result of the non-trivial topology of the {\em bundle} of eigenspaces 
over $\cal M$. By introducing the Wilczek-Zee connection \cite{WIZE}
\begin{equation}
A_{\sigma_i}^{\bar \rho \rho}:= \langle \bar \rho |{\cal U}^\dagger(\sigma)
\,{\partial \over \partial\sigma_i}\,{\cal U}(\sigma)|\rho\rangle \,\, ,
\label{conn}
\end{equation}
where $A_{\sigma_i}^{\bar \rho \rho}$ is the $(\bar \rho, \rho)$ matrix element of the $\sigma_i$ 
component of the connection,
one finds $\Gamma_{A}(C) ={\bf{P}}\exp \int_C A$, \cite{SHWI}, where ${\bf{P}}$ denotes path ordering.
The set Hol$(A):=\{\Gamma_{A}(C);\forall C\in{\cal M}\} \subset U(n)$ is known as the holonomy group
\cite{NAK}. In the case where it coincides with the whole unitary
group $U(n)$ the connection $A$ is called {\em irreducible} \cite{ZARA}.
The transformations $\Gamma_A(C)$ for suitable $C$'s can be used as logical gates for the HQC.

We shall focus on quantum optics, a well established area of quantum physics, in which the developed technology is quite mature as a possible venue for practical implementation of HQC. The model we study here includes laser beams moving through non-linear Kerr media, and acted on by displacing and squeezing devices and interferometers. This implementation has the merit that it gives direct answers to several problems which were raised in the theoretical study of HQC {\cite{PAZARA}.

In Chapter II we present the schematic theoretical description of the quantum optical components employed for HQC. This includes the non-linear Kerr medium, the one and two mode displacing and squeezing devices as well as their effect on the states of laser beams. In Chapter III we construct the non-Abelian Berry connection, the field strength and the holonomies related with this optical setup. A model with $SU(2)$ interferometers is also given as an alternative tool for classical control, and its holonomies are calculated resorting to the non-Abelian Stokes theorem. In Chapter IV the connection between the experimental components of quantum optics and the theoretical requirements for HQC is described. A numerical simulation is finally reported indicating the reliability of the logical gates with respect to the scale resources of the HQC. In the Conclusions the quantum computation characteristics of this model are discussed and issues like the universality, complexity and scalability are addressed.

\section{The Quantum Optical Model}

In the following we shall exploit the advanced tools of quantum optics in order to implement a specific HQC model. All the components used here are thoroughly analyzed in the optics literature \cite{Kral} and experimentally realized by employing such devices as beam splitters, frequency converters, four wave mixers, and others.

\subsection{Kerr medium Hamiltonian and Degenerate States}

In order to perform holonomic computation we shall employ the nonlinear interaction Hamiltonian produced by a Kerr medium
\bq
&&
H_I=\hbar X n(n-1) \,\, ,
\nonumber 
\eq
with $n=a^\dagger a$ the number operator, $a$ and $a^{\dagger}$ being the usual bosonic annihilation and creation operators respectively, and $X$ a constant proportional to the third order nonlinear susceptibility, $\chi^{(3)}$, of the medium. Degenerate eigenstates of $H_I$ are $|0\ran$ and $|1\ran$ ($\{|\nu\ran ; \nu=0,1,... \}$ denoting the Fock basis of number eigenstates $n|\nu\ran=\nu |\nu \ran$). In the case of two laser beams, with annihilation operators $a_1$ and $a_2$ respectively, the total Hamiltonian is given by the sum 
\bq
&&
H_I^{12}=\hbar X n_1(n_1-1) + \hbar X n_2 (n_2-1) \,\, .
\nonumber
\eq
Its degenerate eigenstates are the tensor product of the eigenstates of each subsystem: $|i_1j_2\ran$ $:=|i_1\ran \otimes |j_2 \ran$ for $i_1,j_2=0,1$ with $|i_1\ran$ and $|j_2\ran$ the degenerate states of each beam. Accordingly, the unitary transformations acting on the system are given by the tensor product of the transformations on each individual subsystem. For example, the transformation of a system (Hamiltonian and states) of two lasers when one beam is transformed by $U_1$ is given by the tensor product $U_{12}=U_1\otimes {\bf 1}$. These rules can be applied to build up a system with $m$ lasers. In this case the subspace of Fock states on which we restrict in order to apply the adiabaticity theorem has as basis vectors the degenerate states $|0_l\ran$ and $|1_l\ran$ for each laser labelled by $l$. The general state of the system of $m$ lasers is given by $|\rho_1...\rho_m\ran=|\rho_1\ran \otimes...\otimes |\rho_m\ran$ where $\rho_l$ could be zero or one, for $l=1,...,m$. On this space of states the code can be written. 
We have good reasons to
believe that the problem of the generation of stable Fock states will be
overcome, as suggested by some recent developments \cite{Hong}.  

\subsection{One and Two Laser-Qubit Transformations}

On state $|\psi\ran$ of a laser beam with annihilation operator $a$, the following operators can act
\bq
&&
\text{\it Displacer:} \spp \spp D(\la)=\exp(\la a^\dagger-\bar \la a) \,\, ,
\nonumber
\eq
where $\la$ is an arbitrary complex parameter. The displacing device that implements $D(\la)$ is a simple device that performs a linear amplification to the light field components.
\bq
&&
\text{\it Squeezer:} \spp \spp S(\mu)=\exp(\mu {a^\dagger}^2-\bar \mu a^2) \,\, ,
\nonumber
\eq
where $\mu$ is an arbitrary complex parameter. The squeezing operator $S(\mu)$ can be implemented in the laboratory by a degenerate parametric amplifier.

The transformation operators $D(\la)$ and $S(\mu)$ acting on a single laser beam will result, after a closed loop is performed in their parameter space, into rotations in the state space spanned by $|0\ran$ and $|1\ran$, according to the adiabatic theorem.

The displacer $D(\la)$, transforms the operators $a$, $a^{\dagger}$ and 
any analytic function thereof $f(a,a^{\dagger})$, for any choice of parameters $\lambda$,
as follows \cite{Bishop}
\begin{eqnarray}
&
D(\lambda)aD^{\dagger}(\lambda) = a-\lambda  \,\,\, ,\,\,\,\,\,\,\,\,\, D(\lambda)a^{\dagger}D^{\dagger}(\lambda) = a^{\dagger}-\bar \lambda \,\, ,
&
\no \no
&
D(\lambda)f(a,a^{\dagger})D^{\dagger}(\lambda) =f(a-\lambda,a^{\dagger}-\bar \lambda) \,\, .
&
\nonumber
\end{eqnarray} 
Similarly for the squeezing operator
\begin{eqnarray}
&
S(\mu)aS^{\dagger}(\mu) = \cosh(2r)a+e^{-i \theta} \sinh(2r)a^{\dagger} \,\, ,
&
\no \no
&
S(\mu)a^{\dagger}S^{\dagger}(\mu) = e^{i \theta} \sinh(2r)a+\cosh(2r)a^{\dagger} \,\, ,
&
\no \no
&
S(\mu)f(a,a^{\dagger})S^{\dagger}(\mu) = f\left(S(\mu)aS^{\dagger}(\mu),S(\mu)a^{\dagger}
S^{\dagger}(\mu)\right) \,\, ,
&
\nonumber
\end{eqnarray} 
where $\mu= r e^{i \theta}$, with $r>0$ and $-\pi < \theta \leq \pi$.

On the general state of two lasers $|\psi_{12}\ran=|\psi_1\ran\otimes|\psi_2\ran$ with corresponding annihilation operators $a_1$ and $a_2$, the following operators can act
\bq
&&
\text{\it Two mode squeezer:} \spp \spp M(\zeta)=\exp(\zeta a_1^\dagger a_2^\dagger -\bar \zeta a_1 a_2) \,\, .
\nonumber
\eq
The operator $M(\zeta)$, can be implemented in the laboratory by a non-degenerate parametric amplifier.
\bq
&&
\text{\it Two mode displacer:} \spp \spp N(\xi)=\exp(\xi a_1^\dagger a_2 -\bar \xi a_1 a_2^\dagger) \,\, .
\nonumber
\eq
$M(\zeta)$ and $N(\xi)$ are the transformations between two laser beams that produce, after performing adiabatically a loop in their parametric space, coherent transformations in the two qubit state space spanned by $|00\ran$, $|01\ran$, $|10\ran$ and $|11\ran$. 

These transformations on the states of the laser beams can be produced by $SU(2)$ or $SU(1,1)$ interferometers \cite{Yurke}, according to the algebra which their generators belong to. For instance, each one of $\la a^\dagger-\bar \la a$ and $\xi a_1^\dagger a_2 -\bar \xi a_1 a_2^\dagger$ belongs to an $su(2)$ algebra, while $\mu {a^\dagger}^2-\bar \mu a^2$ and $\zeta a_1^\dagger a_2^\dagger -\bar \zeta a_1 a_2$ belong into (different) $su(1,1)$ algebras \cite{Perelomov}.

\section{Application to The Holonomic Theory}

The non-Abelian Berry connection, $A$, is generated by the topological structure of the bundle of the degenerate sub-spaces. It determines the way to perform a parallel transport of the degenerate eigenstates along an adiabatically spanned loop. In this section we shall show that a complete set of holonomies of $A$ can be {\it explicitly} calculated for our model.

\subsection{The Connection $A$ \label{AAA}}

We initially perform the following polar decomposition of the control variables
\bq
&&
\la=r_0e^{i\th_0}\,\, , \,\,\, \mu=r_1e^{i\th_1} \,\, , \,\,\, \zeta=r_2e^{i\th_2} \,\, , \,\,\, \xi =r_3e^{i\th_3} \,\, .
\nonumber
\eq
We obtain the connection, $A$, from (\ref{conn}), parametrizing the control manifold by the set of real variables introduced above ${\cal M}:=\{r_i,\th_i; \, i=0,...,3\}$ with elements $\sigma_i\in{\cal M}$, where we take ${\cal U}(\sigma)=D(\la)S(\mu)$ for the one laser transformations and ${\cal U}(\sigma)=N(\xi)M(\zeta)$ for transformations between two lasers. We have

\[ \begin{array}{cc}
A_{r_0}=&  
 \left[  \begin{array}{ccc}    0 & -(e^{-i\th_0}\cosh 2r_1-e^{i(\th_0+\th_1)} \sinh 2r_1) \\
   		      		e^{i\th_0}\cosh 2r_1-e^{-i(\th_0+\th_1)} \sinh 2r_1 & 0 \\
\end{array} \right] \,\, , 
\end{array}\]

\[ \begin{array}{cc}
A_{\th_0}=&  
 \left[  \begin{array}{ccc}    ir_0^2 & ir_0(e^{-i\th_0}\cosh 2r_1+e^{i(\th_0+\th_1)} \sinh 2r_1) \\
   		      		ir_0(e^{i\th_0}\cosh 2r_1+e^{-i(\th_0+\th_1)} \sinh 2r_1)& ir_0^2 \\
\end{array} \right] \,\, .
\end{array}\]
For the connection components, $A_{r_0}$ and $A_{\th_0}$, it is more convenient to use for the variables $\la$ the decomposition $\la=x+iy$, with $x$ and $y$ real, resulting into the following components of the connection
\[ \begin{array}{cc}
A_x=\cos\th_0 A_{r_0}-{\sin \th_0 \over r_0}A_{\th_0}=&  
 \left[  \begin{array}{ccc}     -iy & -(\cosh 2r_1 - e^{i\th_1} \sinh 2r_1) \\
   		      		\cosh 2r_1-e^{-i\th_1} \sinh 2r_1& -iy \\
\end{array} \right] \,\, ,
\end{array}\]

\[ \begin{array}{cc}
A_y=\sin \th_0 A_{r_0} + {\cos \th_0 \over r_0} A_{\th_0}=&  
 \left[  \begin{array}{ccc}     ix & i(\cosh 2r_1 + e^{i\th_1} \sinh 2r_1) \\
   		      		i(\cosh 2r_1 + e^{-i\th_1} \sinh 2r_1) & ix \\
\end{array} \right] \,\, ,
\end{array}\]

\[ \begin{array}{ccc}
A_{r_1}=&  
 \left[  \begin{array}{ccc}    0 & 0 \\
   		      	       0 & 0 \\
\end{array} \right] \,\,\,\,\,\, ,\,\,\,\,\,\,\,\,\,\,\,\,\,\,
%
A_{\th_1}=&  
 \left[  \begin{array}{ccc}    1 & 0 \\
   		      	       0 & 3 \\
\end{array} \right] {i \over 4} (\cosh 4 r_1 -1)  \,\, ,
\end{array}\]

\[ \begin{array}{ccc}
A_{r_2}=&  
 \left[  \begin{array}{cccc}    0 & 0 & 0 & -e^{-i\th_2}\\
   		      	        0 & 0 & 0 & 0 \\
				0 & 0 & 0 & 0 \\
				e^{i\th_2} & 0 & 0 & 0 \\    
\end{array} \right] \,\,\,\,\,\, ,\,\,\,\,\,\,\,\,\,\,\,\,\,\,
A_{r_3}=&  
 \left[  \begin{array}{cccc}    0 & 0 & 0 & 0 \\
   		      	        0 & 0 & -e^{-i\th_3} & 0 \\
				0 & e^{i\th_3} & 0 & 0 \\
				0 & 0 & 0 & 0 \\    
\end{array} \right] (2 \cosh ^2 r_2 -1) \,\, .
\end{array}\]
The components $A_{\th_2}$ and $A_{\th_3}$ have more complicated forms that we shall not give explicitly here as they are not necessary for performing universal quantum computation \cite{UG}.

\subsection{The Commutators and the Field Strengths $F$\label{comm}}

In order to be able to calculate the holonomies \cite{PAZARA} it is convenient to consider loops $C$ on the planes $(\sigma_i,\sigma_j)$
in ${\cal M}$,\footnote{Note that the components ($r_0$,$\th_0$) have been 
replaced by ($x$,$y$).} on which the two components of the connection commute with each other 
yet giving a non-trivial holonomy (i.e. they have non-zero field strength component, $F_{\sigma_i \sigma_j}=\p_{\sigma_i} 
A_{\sigma_j}-\p_{\sigma_j}A_{\sigma_i}+[A_{\sigma_i},A_{\sigma_j}]$).
Indeed, for $\hat \sigma_i$, $i=1,2,3$, denoting the Pauli matrices, we have
\bq
&
[A_x,A_{r_1}]=0\,\,\,\,\, {\text{with}} \,\,\,\, 
\left. F_{xr_1} \right|_{\th_1=0}=-2 i {\hat \sigma}_2 \exp(-2 r_1) \,\, ,
&
\no \no
&
[A_y,A_{r_1}]=0\,\,\,\,\, {\text{with}} \,\,\,\, 
\left. F_{yr_1} \right|_{\th_1=0}=-2 i {\hat \sigma}_1 \exp(2 r_1) \,\, ,
&
\no \no
&
[A_{r_1},A_{\th_1}]=0\,\,\,\,\, {\text{with}} \,\,\,\, 
F_{r_1 \th_1}= - i \hat{s}_3 \sinh 4 r_1 \,\, ,
&
\no \no
&
[A_{r_2},A_{r_3}]=0\,\,\,\,\, {\text{with}} \,\,\,\, 
\left. F_{r_2 r_3} \right|_{\th_2=\th_3=0}=-2i \hat \sigma_2^{12}\sinh 2 r_2 \,\, ,
&
\no \no
&
[A_{r_2},A_{r_3}]=0\,\,\,\,\, {\text{with}} \,\,\,\, 
\left. F_{r_2 r_3} \right|_{\th_2=0, \th_3=3\pi/2}=-2i \hat{\sigma}_1^{12} \sinh 2 r_2 \,\, ,
&
\nonumber
\eq
where
\bq
&&
\hat{s}_3:= - \left[  \begin{array}{cccc}    1 & 0 \\
  	         		      	        0 & 3 \\
\end{array} \right]
\Sp , \Sp
\hat \sigma _2^{12}:=
\left[  \begin{array}{cccc}    0 & 0 & 0 & 0 \\
   		      	        0 & 0 & -i & 0 \\
				0 & i & 0 & 0 \\
				0 & 0 & 0 & 0 \\
\end{array} \right]
\Sp \text{and} \Sp
\hat \sigma _1^{12}:=
\left[  \begin{array}{cccc}    0 & 0 & 0 & 0 \\
   		      	        0 & 0 & 1 & 0 \\
				0 & 1 & 0 & 0 \\
				0 & 0 & 0 & 0 \\
\end{array} \right] \,\, .
\nonumber
\eq
The above conditions that are satisfied on the planes $\left.(x,r_1)\right._{\th_1=0}$, $\left.(y,r_1)\right._{\th_1=0}$, $(r_1,\th_1)$, $\left.(r_2,r_3)\right._{\th_2=\th_3=0}$ and $\left.(r_2,r_3)\right._{\th_2=0, \th_3=3\pi/2}$ allow for the explicit calculation of the holonomies for paths restricted on such planes. 

\subsection{The Holonomies $\Gamma_A(C)$}

In order to perform universal quantum computation it is necessary to produce at least two independent unitary gates \cite{Lloyd}. In the following we shall present holonomic gates, which involve (any) one qubit rotations and a special class of (any) two qubit transformations. In detail we have
\bq
&
C_I \in \left. (x,r_1)\right._{\th_1=0} \,\,\,\text{gives}\,\,\, \Gamma_A(C_I)=\exp -i\hat \sigma_1 \Sigma_I \,\,\, \text{with} \,\,\, \Sigma_I:=\int_{\Sigma(C_I)} \!dxdr_1 2 e^{-2r_1} \,\, ,
&
\no \no
&
C_{II} \in \left. (y,r_1)\right._{\th_1=0} \,\,\, \text{gives} \,\,\, \Gamma_A(C_{II})=\exp -i\hat \sigma_2 \Sigma_{II} \,\,\, \text{with} \,\,\, \Sigma_{II}:=\int_{\Sigma(C_{II})} \! dydr_1 2 e^{2r_1} \,\, ,
&
\no \no
&
C_{III} \in (r_1,\th_1) \,\,\, \text{gives} \,\,\, \Gamma_A(C_{III})=\exp -i\hat{\tilde{\sigma}}_3 \Sigma_{III} \,\,\, \text{with} \,\,\, \Sigma_{III}:=\int_{\Sigma(C_{III})} \! dr_1d\th_1 \sinh 4r_1 \,\, ,
&
\no \no
&
C_{IV} \in \left. (r_2,r_3)\right._{\th_2=\th_3=0} \,\,\,\text{gives} \,\,\,\Gamma_A(C_{IV})=\exp -i\hat \sigma_2^{12} \Sigma_{IV} \,\,\, \text{with} \,\,\, \Sigma_{IV}:=\int_{\Sigma(C_{IV})} \! dr_2dr_3 2 \sinh 2 r_2 \,\, ,
&
\no \no
&
C_{V} \in \left. (r_2,r_3)\right._{\th_2=0, \th_3=3\pi/2} \,\,\,\text{gives} \,\,\, \Gamma_A(C_{V})=\exp -i\hat \sigma_1^{12} \Sigma_{V} \,\,\, \text{with} \,\,\, \Sigma_{V}:=\int_{\Sigma(C_{V})} \! dr_2dr_3 2 \sinh 2 r_2 \,\, ,
&
\no \nonumber
\eq
where $\Sigma(C_\rho)$ with $\rho=I,...,V$ is the surface on the relevant submanifold $(\sigma_i,\sigma_j)$ of ${\cal M}$ whose boundary is the path $C_\rho$. The hyperbolic functions in these integrals stem out of the geometry of the $su(1,1)$ manifold associated with the relative control submanifold. The $\Gamma_A(C)$'s thus generated belong either in the $U(2)$ or $U(4)$ group. Considering the tensor product structure of our system these rotations represent in the $2^m$ space of $m$ qubits respectively single qubit rotations and two qubit interactions, thus resulting into a universal set of logical gates. Their explicit constructions are similar to those presented in \cite{PAZARA} for the ${\bf CP}^n$ model.

\subsection{The $SU(2)$ Control Manifold \label{su2}}

In what follows we discuss the employment of $SU(2)$ interferometer as control devices \cite{Yurke} for producing holonomies. For $a_1$ and $a_2$ the annihilation operator of two different laser beams, consider the Hermitian operators
\be
J_x={1 \over 2} (a_1^{\dagger} a_2 +a_2^\dagger a_1)
\,\, , \,\,\,
J_y=-{i \over 2} (a_1^{\dagger} a_2 -a_2^\dagger a_1)
\,\, , \,\,\,
J_z={1 \over 2} (a_1^{\dagger} a_1 -a_2^\dagger a_2)
\label{JJJ}
\ee
and
\bq
&&
N=a_1^{\dagger} a_1 +a_2^\dagger a_2=n_1+n_2 \,\, .
\nonumber
\eq
The operators (\ref{JJJ}) satisfy the commutation relations for the Lie algebra of $SU(2)$; $[J_x,J_y]=iJ_z$, $[J_y,J_z]=iJ_x$, $[J_z,J_x]=iJ_y$. The operator $N$, which is proportional to the free Hamiltonian of two laser beams, commutes with all of the $J$'s. On the other hand, however, the Kerr Hamiltonian does not commute with the $J$'s, allowing for the possibility that $SU(2)$ interferometers be used as transformation controllers in view of the holonomic computation. 

From these operators we obtain the unitaries, $U_x(\al)=\exp(i\al J_x)$, $U_y(\beta)= \exp (i \beta J_y)$ and $U_z(\gamma)=\exp(i\gamma J_z)$. For the degenerate state space of two laser beams spanned by $|i_1 j_2\ran$, we have from (\ref{conn}) and for ${\cal U}=U_x(\al)U_y(\beta)U_z(\gamma)$ the following connection components
\be
A_\al ={i \over 2}
 \left[  \begin{array}{cccc}    0 & 0 & 0 & 0 \\
				0 & \sin \beta & \cos \beta e^{i\gamma} & 0 \\
				0 & \cos \beta e^{-i \gamma} & -\sin \beta & 0 \\
				0 & 0 & 0 & 0 \\
\end{array} \right] 
,\,\,\,
A_\beta=-{1 \over 2}
 \left[  \begin{array}{cccc}    0 & 0 & 0 & 0 \\
				0 & 0 & e^{i\gamma} & 0 \\
				0 & -e^{-i \gamma} & 0 & 0 \\
				0 & 0 & 0 & 0 \\
\end{array} \right] 
,\,\,\,
A_\gamma=-{i \over 2}
 \left[  \begin{array}{cccc}    0 & 0 & 0 & 0 \\
				0 & 1 & 0 & 0 \\
				0 & 0 & -1 & 0 \\
				0 & 0 & 0 & 0 \\
\end{array} \right] .
\label{conn1}
\ee
These components do not commute with each other, when projected on planes with non-trivial field strength. Hence, it is not possible to employ again the method used in the previous section to calculate the holonomies of paths in the three dimensional control parameter space, $(\al, \beta,\gamma)$. Instead, for this purpose we may employ the non-Abelian Stokes theorem \cite{Karp}. The extra limitation, now, for the choice of the path comes from the constraint that, apart from being confined on a special two dimensional subspace, it has to have the shape of an orthogonal parallelogram with two sides lying along the coordinate axis. This will facilitate the extraction of an analytic result from the Stokes theorem. Though, experimentally, this restriction poses additional (possibly minor) difficulties, theoretically it leads to the very interesting possibility of a direct calculation of non-Abelian holonomies without resorting to their Abelian substructures. Note that the application of the Stokes theorem for the evaluation of the holonomies in the previous section gives the same results, as expected.

To state the non-Abelian Stokes theorem let us first present some preliminaries, where a few simplifications are introduced, as its general form will not be necessary in the present work. 

\begin{figure}[h]
  \epsffile{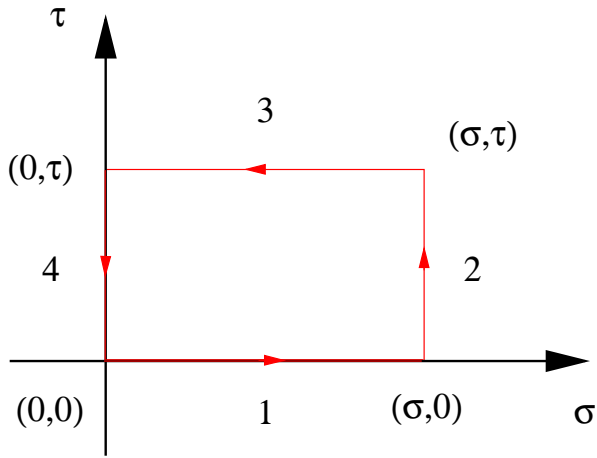}
  \caption[contour]{\label{loop}
The loop $C$ for the non-Abelian Stokes theorem.
           }
\end{figure}

Consider the Wilson loop (holonomy), $W={\bf P}\exp \oint_C A$, of the loop $C$ given in Fig. \ref{loop}, with connection $A$, made out of the Wilson lines $W_i$ for $i=1,...,4$, as $W=W_4 W_3 W_2 W_1$. $(\sigma,\tau)$ is a reparametrization of the plane where the loop $C$ lies. Define $T^{-1}(\sigma,\tau)=W_4 W_3$. Then, for $F_{\sigma\tau}$ the field strength of the connection $A$ on the plane $(\sigma,\tau)$, $W$ is given in terms of a surface integral
\bq
&&
W={\bf P}_\tau e^{\int_\Sigma T^{-1}(\sigma,\tau)F_{\sigma \tau}(\sigma,\tau)T(\sigma,\tau) d\sigma d\tau} \,\, ,
\nonumber
\eq
where ${\bf P}_\tau$ is the path ordered symbol with respect only to the $\tau$ variable, contrary to the usual path ordering symbol {\bf P}, which is with respect to both variables, $\sigma$ and $\tau$. Here $F_{\sigma \tau}(\sigma,\tau)=-\p_\sigma A_\tau +\p_\tau A_\sigma+[A_\sigma , A_\tau]$.

From the connection given in (\ref{conn1}) the following holonomies are derived. For a closed rectangular loop $C_1\in (\al,\beta)$-plane with coordinates $(0,0),(\al \! = \! \pi,0),(\al \! = \! \pi,\beta),(0,\beta)$ we obtain the following unitary transformation
\be
\Gamma_A(C_1)=\exp (-i2\beta \hat \sigma_2^{12}) \,\, .
\label{C1}
\ee
In addition for a rectangular loop $C_2\in (\al,\gamma)$-plane with coordinates $(0,0),(\al \! = \! \pi,0),(\al \! = \! \pi,\gamma),(0,\gamma)$ we obtain the holonomy
\be
\Gamma_A(C_2)=\exp (-i2\gamma \hat \sigma_3^{12}) \,\, ,
\ee
where the matrix $\hat \sigma_3^{12}$ is defined similarly to $\hat \sigma_1^{12}$ and $\hat \sigma_2^{12}$ in Subsection \ref{comm}. These operations can be implemented by using interferometers between {\it any} two laser beams. Note that the coefficients in front of the matrices in the unitaries are areas on spheres spanned by the angles $\al$ and $\beta$ or $\al$ and $\gamma$. This is consistent with the geometry of $SU(2)$. 

These two matrices can produce any unitary transformation of {\it one} qubit encoded in a sub-space of states of the two laser beams spanned by $\{|01\ran,|10\ran\}$. In other words, we need two laser beams to encode one qubit, contrary to previous construction. As these transformations can be performed between any two beams, we can generate interaction transformations between two qubits, resulting finally (together with the one qubit rotations) into a universal set of transformations. For example the SWAP two qubit gate given by
\bq
&&
U_{SWAP}= \left[  \begin{array}{cccc}   1 & 0 & 0 & 0 \\
			 		0 & 0 & 1 & 0 \\
					0 & 1 & 0 & 0 \\
					0 & 0 & 0 & 1 \\
\end{array} \right] \,\, ,
\nonumber
\eq
is achieved as follows. On four arbitrary laser beams $1,2,3$ and $4$ with $\{|01\ran,|10\ran\}_{1,2}$ encoding the one qubit and $\{|01\ran,|10\ran\}_{3,4}$ encoding the other, we may act with $\left.\Gamma_A(C_1)\right|_{\beta={\pi \over 4}}$ between beams $1$ and $3$ and with $\left.\Gamma_A(C_1)\right|_{\beta={3 \pi \over 4}}$ between $2$ and $4$ producing eventually the $U_{SWAP}$ gate. The loop $C_1$ is defined as in (\ref{C1}).

This model facilitates the physical implementation as it will be seen in the following section.

\section{Towards experimental implementation}

We address here the task of combining the theoretical requirements of HQC together with the features of the ``experimental'' components described in the previous two sections. While the Abelian holonomies have been produced in the laboratory by various means, the non-Abelian ones are more complicated. However, the holonomies calculated above, require successive restrictions on two dimensional planes of the control parameter space, quite in the same way as one needs to do to generate Abelian Berry phases. This constructive method may prove experimentally advantageous for performing and measuring non-Abelian holonomies. A survey over some Berry phase experiments in optics is given below.

\subsection{Various Abelian Berry Phase Setups}

Photons can be seen as massless spin-1 bosons. This characteristic has been the
driving force 
for the optical manifestation of the Berry phase with respect to the polarization 
quantum numbers \cite{Simon}.
Necessary condition for the generation of this phase factor is the
adiabatic
change of the direction of the photon propagation. Various optical experiments
have been performed. Results at the classical level have been reported in \cite{Chiao}, for the
case of a single mode in a wounded optical fiber, whereas quantum mechanically, in \cite{Kwiat},
the case of a single photon has been treated.
Of special importance, for our case, is the latter experiment where the
Berry phase has been observed
at quantum optical level. In this case the incident
light is prepared in an entanglent state 
\bq
&&
|\psi\rangle_{in}  = \int A(E')|n \rangle_{E'}   |n\rangle_{E-E'} dE' \,\, ,
\nonumber
\eq
where $A(E')=A(E-E')$ is the complex probability amplitude for finding one
photon with an 
energy $E'$ ($|n =1\rangle_{E'}$) or with an energy $E-E'$ ($|n=1\rangle_{E-E'}$).
This type of states can be 
produced in the lab by driving a single-mode ultraviolet laser
into a $\chi ^{(2)}$ nonlinear optical crystal. A Michelson interferometer
has been used for the observation of the phase in the output state. It was
found that the output state (photons in essentially $n$=1 Fock states) had
an extra phase factor due to the optical-path-length difference $\Delta L$
of the interferometer plus the contribution of the Berry phase. 
The form of such state is given by
\bq
&&
|\psi\rangle_{out}  = \frac{1}{\sqrt{2}} \int A(E')| n\rangle_{E'}
|n\rangle_{E-E'} \{1+\exp[i\phi (E-E')]\} dE' \,\, ,
\nonumber
\eq
where $\phi (E-E') = 2\pi\Delta L / \lambda_{E-E'} + \phi_{Berry}$, with $\phi_{Berry}$ the geometrical phase predicted theoretically.

Recently, an alternative approach to the geometric phase 
has been considered \cite{Jackiw}, through  squeezed
states of photons. Squeezed states have been 
found considerably interesting in the field of quantum optics for 
various reasons, as for example, the noise reduction which is
necessary for practical applications with noise sensitivity.  
   
Displacement and squeezing give different contributions to the Berry
phase of the Fock states $|\nu\ran$.
For the case of squeezing one finds that this contribution is given 
by
\bq
&&
\phi_{Berry}^n  = \frac{2n+1}{4} \oint \left( \cosh 4r_1 -1 \right)dr_1 \,\, .
\nonumber
\eq
Such Berry phase agrees with the form of the diagonal connection $A_{\theta_1}$ in Subsection \ref{AAA}, as it was to be expected. On the other hand if we perform a loop in the control parameters of the displacing device we expect the following Berry phase to arise
\bq
&&
\phi_{Berry}^n = \oint (ydx-xdy) \,\, .
\nonumber
\eq
The equivalent connection of displacing in the Kerr medium ($A_x$ and $A_y$ with $r_1=0$, in Subsection \ref{AAA}) are non-diagonal matrices, whose holonomy cannot be calculated easily. In fact, as it is observed by the numerical simulations in the following, the phase factors produced in front of $|0\ran$ and $|1\ran$ are not equal, due to the off-diagonal elements of $A_x$ and $A_y$. This effect is related with the degeneracy structure of each model.

\subsection{Free Hamiltonian and Kerr Medium}

In the previous sections we used the Kerr non-linear Hamiltonian in order to produce the degenerate eigenspace spanned by $|0\ran$ and $|1\ran$. The full Hamiltonian of the system is the combined one of the free photons and the non-linear medium, i.e. $H_{Tot}=H_{Free}+H_{Kerr}=\hbar \omega n +\hbar X n(n-1)$. Of course the first part lifts the degeneracy of $|0\ran$ and $|1\ran$ destroying the basic requirements for the holonomic computation. In order to overcome this problem we resort to the following constructions.

Considering $H_{Free}$ as unperturbed Hamiltonian and the non-linear part as the interaction term we may move to the interaction picture of the full system, with $H_I= H_{Kerr}$. The rotation to the interaction picture may be incorporated in the devices used for the external control resulting in a redefinition of their control parameters.

Alternatively, we may define a one dimensional lattice with points on the trajectory of the laser. As the free Hamiltonian is acting only on the state $|1\ran$ changing its phase by $e^{-iH_{Free} \Delta t}=e^{-i \hbar \omega \Delta x /c}$, with $c$ the speed of light, we may single out the points $x_k =2\pi c k /(\hbar \omega)$ for $k$ integers. On these points the phase is trivial and it does not contribute to the state. Hence, $|0\ran$ and $|1\ran$ are degenerate on this lattice \cite{Kitano}. 

In Subsection \ref{su2} we have introduced $SU(2)$ interferometers as control devices. The $su(2)$ operators commute with $H_{Free}$ allowing the effect of the free Hamiltonian to factorize out of the whole control procedure. At the end of the algorithm the detectors may be placed on a point of the degenerate lattice in order to avoid the dynamical phase produced by $H_{Free}$ on the states $|0\ran$ and $|1\ran$. Even though in the $SU(2)$ model each qubit is encoded with the help of two laser beams increasing in this way the necessary resources, it overcomes the problem of the degeneracy in the most efficient way.

\subsection{Holonomies and Devices}

For the implementation of the continuous adiabatic loops we should adopt the kick method described in \cite{PAZARA}, \cite{VIO} and \cite{Vitali}. A general state $|\psi\ran$ in the degenerate eigenspace of $H_0=H_{Kerr}$ is given as a linear combination of $|0\ran$ and $|1\ran$. Under an isospectral cyclic evolution of the Hamiltonian in the family ${\cal F}$, the evolution operator acting on $|\psi\ran$ is given by the $2 \times 2$ submatrix in the upper left corner of
\bq
&&
U(0,T)={\bf T} \exp -i \int ^T_0 {\cal U}(\sigma (t))H_0 {\cal U}^\dagger(\sigma(t))dt \,\, .
\nonumber
\eq
This evolution takes place from time $0$ to time $T$ and, for performing a closed loop, we demand $\sigma(0)=\sigma(T)$. By dividing the time interval, $[0,T]$, into $m$ equal segments $\Delta t$ we may approximate the above operator by
\bq
&&
U(0,T)\approx{\bf T} \prod_{i=1}^{m} {\cal U}_ie^{-iH_0 \Delta t} {\cal U}_i^\dagger \,\,\,\,\,\,\,\, \text{with} \,\,\,\, {\cal U}_i={\cal U}(\sigma_i)={\cal U}(\sigma(t_i)) \,\, .
\nonumber
\eq
Assuming the evolutions ${\cal U}_{i+1}^{\dagger}{\cal U}_i$ to be a very small rotation and restricting to evolutions which remain in the zeroth degenerate eigenspace we might once more derive the holonomy operator $\Gamma_A(C)$ for $A$ defined in (\ref{conn}). We prefer instead to see what the effect of finitely many devices would be, when acting on the space of states of the qubits (the lasers). 

For the sake of concreteness we work out examples in terms of displacing devices $D(\la)$, performing a closed loop in their control parameters $\la$. This is shown in Fig. \ref{trig}, where for simplicity the least possible number of displacing devices (three) for performing a closed loop has been considered. Two displacing unitaries are combined as $D(\lambda)D(\lambda')=\exp{(i \Im (\lambda \bar \lambda'))} D(\la+\la')$. The physical process behind this is as follows. On the state $|\psi\ran$ first acts a displacing device with unitary $D^\dagger(\la_1)$, taking it to the point $\la_1$. Then, the evolution operator of the Kerr Hamiltonian acts for a time interval $\Delta t= T/3$ $U(\Delta t)=\exp( -i H_0 \Delta t)$. This effect is achieved by propagating the beam inside a Kerr medium. Then, the evolution $D^{\dagger}(\la_2)D(\la_1)$ is performed. This is achieved, with a single displacing device, given (up to an overall phase factor that will cancel at the end) by $D(\la_1-\la_2)$. After exiting the displacing device (we are at point $\la_2$) the beam enters a Kerr medium for time $\Delta t$ and then the procedure is repeated until we come back to the point $\la_1$ and the beam enters once more the Kerr medium. Finally, the state is thus displaced by $D(\la_1)$. This loop may be transported to any other place of the control parameter complex plane by acting at the beginning and at the end of this procedure with the appropriate displacing unitary (device).
\begin{figure}[h]
  \epsffile{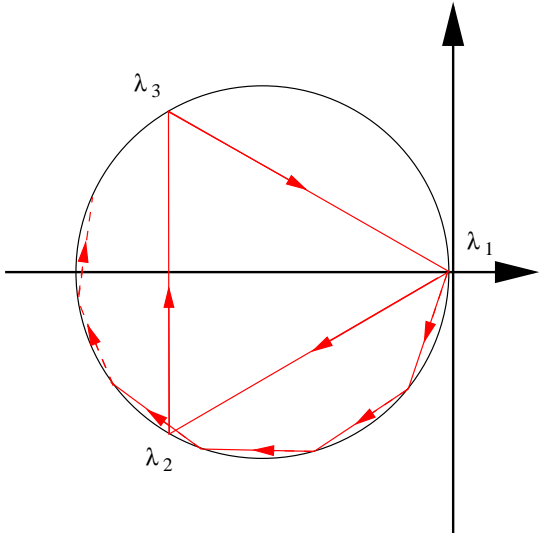}
  \caption[contour]{\label{trig}
The triangular (and polygonal) loop $C$ on the complex plane of the displacing control parameters, $\la$, approximating the circle.
           }
\end{figure}
In this case the evolution operator is approximated by 
\bq
&&
U(0,T)\approx D(\la_1) \left( U(\Delta t;0) U(\Delta t;\tilde \la_1 +\tilde \la_2 ) U(\Delta t; \tilde \la_1) U(\Delta t;0) \right) D^\dagger(\la_1) \,\, ,
\nonumber
\eq
where $U(\Delta t;\tilde \la) =D(\tilde \la) U(\Delta t) D^\dagger(\tilde \la)$, $\tilde \la_i=\la_{i+1}-\la_i$ and $\la_4=\la_1$.

According to the above analysis, we proceed to the numerical simulation of a system with various numbers of displacing devices represented by different polygons on the control complex plane (see Fig. \ref{trig}). We start with a pentagon which demands five displacers. Fig. \ref{sub} represents the absolute values of the (0,0), (0,1), (1,0), (1,1) elements of the evolution operator $U(0,T)$ as functions of the number of displacers used to approximate a cyclic evolution. These are the relevant elements for the evolution of the states in the degenerate eigenspace describing a qubit. The parameters involved are taken to be $T=0.1$ and $\hbar X=1$, with the radius of the circle equal to 1. The initial point is taken to be the origin of the complex plane rather than $\la_1$, or in other words we do not perform the initial and final displacings by $D^\dagger(\la_1)$ and $D(\la_1)$.
\begin{figure}[h]
  \epsffile{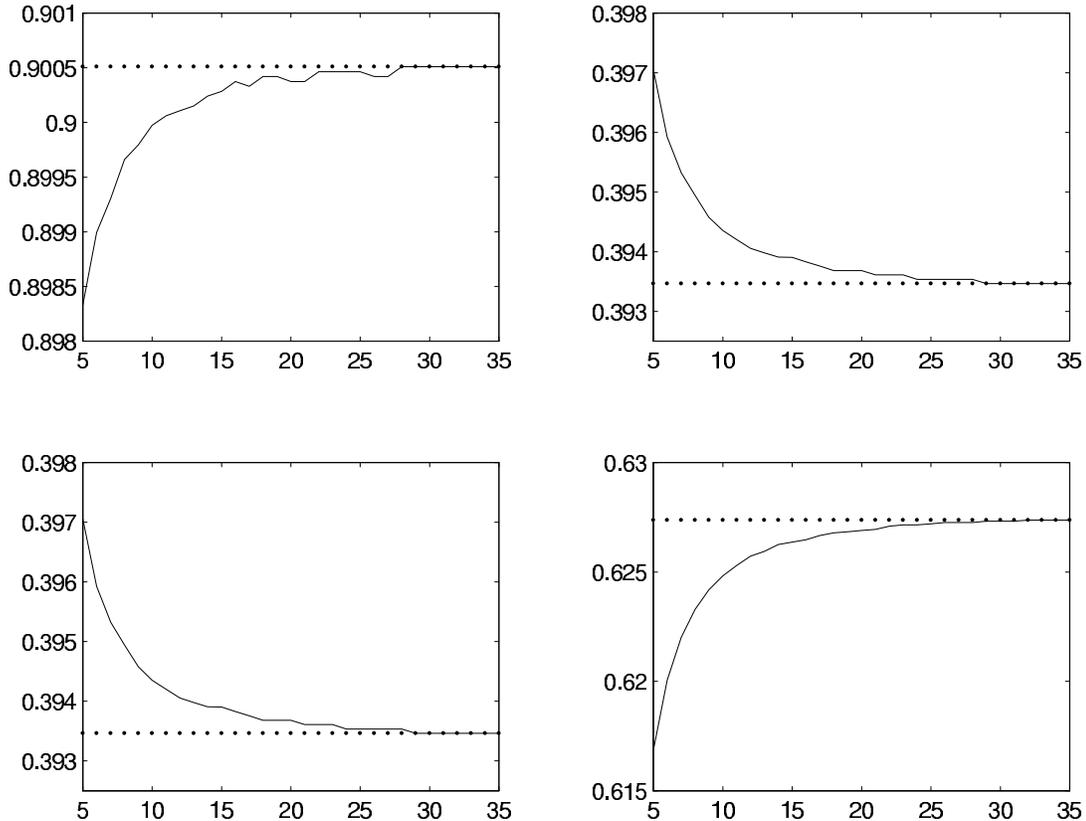}
  \caption[contour]{\label{sub}
The absolute values $|U_{00}(0,T)|$, $|U_{01}(0,T)|$, $|U_{10}(0,T)|$ and $|U_{11}(0,T)|$ as functions of the number of sides of the polygons.
           }
\end{figure}
In the table below are depicted the percent deviations of those values obtained with 5, 10, 20 and 26 displacers with respect to the ones obtained with 100 displacers.
\begin{center}
\begin{tabular}{|c||c|c|c|c|}                   \hline
   &  5     & 10     & 20     & 26     \\       \hline \hline
\Sp 00 \Sp & \Sp 0.2419 \% \Sp & \Sp 0.0595 \% \Sp & \Sp 0.0149 \% \Sp & \Sp 0.0099 \% \Sp \\       \hline
\Sp 01 \Sp & 0.9119 \% & 0.2260 \% & 0.0558 \% & 0.0186 \% \\       \hline
\Sp 10 \Sp & 0.9119 \% & 0.2260 \% & 0.0558 \% & 0.0186 \% \\       \hline
\Sp 11 \Sp & 1.6763 \% & 0.4061 \% & 0.0760 \% & 0.0269 \% \\       \hline
\end{tabular}
\end{center}
We see that with 26 displacers the error is of the order of 1 in $10^4$ acceptable for quantum computation with error correction. This provides an indication for the necessary number of devices needed in order to reproduce faithfully the holonomic adiabatic loop. 

\section{Conclusions}

The implementation of HQC in the frame of quantum optics has provided novel insight into many technical aspects of the theory. Moreover, the components demanded for it are widely used in the laboratories. The possibility of overcoming the difficulties in combining them in the appropriate way for obtaining holonomies is an open problem to be faced by experimenters. 

In summary the main quantum computational features we observed in our model are the following. First, the {\it universality} condition is proven explicitly, stemming out of the ability to construct holonomies representing any possible logical gate. This is achieved by combining one qubit rotations (realized by displacing and squeezing devices) and two qubit transformations (by interferometers) between any two qubits. Second, the setup exhibits quantum entanglement, having built in tensor product structure as it consists of a multi partite system. This resolves the problem of {\it complexity} posed in \cite{PAZARA} which is one of the main features which make QC's more efficient than classical ones. Third, the degenerate space of the Hamiltonian eigenstates, which is used to write the code is constructed out of laser beams each with a two dimensional degenerate space. So the demand of using a big degenerate space to write useful codes is performed not by resorting to one system with very large degeneracy, which is almost impossible to realize in nature, but by adding up the 2-dimensional subspaces of the lasers. This is the characteristic of {\it scalability} of the proposed model. Fourth, the chosen loops associated with the given holonomies are restricted on specific planes $(\sigma_i,\sigma_j)$ of two control parameters $\sigma_i$ and $\sigma_j$, exactly in the same way as used for the production of {\it Abelian Berry phases}. The latter has been verified in several theoretical and experimental applications in optics \cite{Simon,Chiao,Kwiat} and elsewhere \cite{ELSE}. From these phase transformations $U(1)$ of different components of the system we are able to obtain with proper combinations any desired $U(2^m)$ transformation. Since there exist experimental measurements of the Berry phase, it is plausible to expect the implementation of the $U(2^m)$ holonomic transformations.

A further final advantage of the holonomic setup is that it is confined in the degenerate eigenspace produced by $\{|0\ran,|1\ran\}$, describing one qubit. Entanglement of these states with the non-degenerate ones in the course of application of the logical gates does not occur due to the adiabaticity requirement. The initial control operators we use here, $D$, $S$, $M$ and $N$ in general mix all the states of the Fock space, but at the end of the loop, only rotations between the degenerate eigenstates will be accounted for.

The possibility to observe the proposed holonomies in the laboratory or even perform specific logical gates is a demanding task and an open question for the future.

\section{Acknowledgements}

We would like to thank Mario Rasetti, Paolo Zanardi and Matteo Paris for inspiring conversations. This work was supported in parts by TMR Network under the condract no. ERBFMRXCT96 - 0087.


\begin{references}

\bibitem{ZARA} P. Zanardi and M. Rasetti, to appear in Phys. Lett. A, quant-ph/9904011. For related works see 
A. Kitaev, quant-ph/9707021; J. A. Jones, V. Vedral, A. Ekert and G. Castagnoli, quant-ph/9910052;
K. Fujii, quant-ph/9910069.

\bibitem{PAZARA} J. Pachos, P. Zanardi and M. Rasetti, 
to appear in Phys. Rev. A (Rapid Comm.), quant-ph/9907103.

\bibitem{QC} For reviews, see D.P. DiVincenzo, {\sl Science} {\bf
270}, 255 (1995); A. Steane, Rep. Prog. Phys. {\bf 61}, 117 (1998).

\bibitem{Chuang} I. L. Chuang and Y. Yamamoto, quant-ph/9505011.

\bibitem{SHWI} For a review see, {\em Geometric Phases in Physics}, A. Shapere and F. Wilczek, Eds.
World Scientific (1989).

\bibitem{WIZE} F. Wilczek and A. Zee, Phys. Rev. Lett. {\bf {52}}, 2111 (1984).

\bibitem{NAK} M. Nakahara, {\em Geometry, Topology and Physics}, IOP Publishing Ltd. (1990).

\bibitem{Kral} V. Buzek and P.L Knight, in Progress in Optics XXXIV, E. Wolf (North
Holland, Amsterdam) (1995); P. Kral, Phys. Rev. A, {\bf 42}, 4177 (1990), J. Mod. Opt. {\bf 37}, 889 (1990);
C. F. Lo, Phys. Rev. A {\bf 43}, 404 (1991); M. G. A. Paris, Phys. Lett. A, {\bf 217}, 78 (1996).

\bibitem{Hong} J. I. Cirac, R. Blatt, A. S. Parkins and P. Zoller, Phys. Rev. Lett., {\bf 70}, 762 (1993);
T. Pellizzari and H. Ritsch, Phys. Rev. Lett., {\bf 72}, 3973 (1994);
T. Pellizzari and H. Ritsch, Phys. Rev. Lett., {\bf 72}, 3973 (1994);
M. G. A. Paris, M. B. Plenio, S. Bose, D. Jonathan and G. M. D'Ariano,
quant-ph/9911036.

\bibitem{Bishop} R. F. Bishop and A. Vourdas, J. Phys. A, {\bf 20}, 3743 (1987), Phys. Rev. A, {\bf 50}, 4488 (1994).

\bibitem{Yurke} B. Yurke, S. L McCall and J. R. Klauder, Phys. Rev. A, {\bf 33}, 4033 (1986);
C. Brif and A. Mann, Phys. Rev. A, {\bf 54}, 4505 (1996);
C. Brif and Y. Ben-Aryeh, Quant. Semiclass. Opt., {\bf 8}, 1 (1996).

\bibitem{Perelomov} A. Perelomov, {\em Generalized Coherent States and their Applications}, Springer-Verlag (1986).

\bibitem{UG} D. Deutsch, A. Barenco and A. Ekert, Proc. R. Soc. London { A}, {\bf 449}, 669 (1995);
D.P. Di Vincenzo, Phys. Rev. A, {\bf 50}, 1015 (1995).

\bibitem{Lloyd} S. Lloyd, Phys. Rev. Lett., {\bf 75}, 346 (1995).

\bibitem{Karp}  R. Karp, F. Mansouri and J. Rno, to appear in Jour. Math. Phys., hep-th/9910173.

\bibitem{Simon} A. Simon, Phys. Rev. Lett., {\bf 51}, 2167 (1983); 
J. N. Ross, Opt. Quantum Electron. {\bf 16}, 455 (1984); 
P. Facchi and S. Pascazio, submitted to Acta Physica Slovaca, quant-ph/9904082.

\bibitem{Chiao} R. Y. Chiao and Y-S. Wu, Phys. Rev. Lett., {\bf 57}, 933 (1986); 
A. Tomita and R. Y. Chiao, Phys. Rev. Lett., {\bf 57}, 937 (1986).

\bibitem{Kwiat} P. G. Kwiat and R. Y. Chiao, Phys. Rev. Lett., {\bf 66}, 588 (1991).

\bibitem{Jackiw} R. Jackiw and A. Kerman, Phys. Lett., A, {\bf 71}, 158 (1979); 
J. Liu, B. Hu and B. Li, cond-mat/9808084; 
S. Seshadri, S. Lakshmibala and V. Balakrishnan, quant-ph/9905101.

\bibitem{Kitano} M. Kitano, quant-ph/9505024.

\bibitem{VIO} L. Viola, E. Knill and S. Lloyd, Phys. Rev. Lett., {\bf 82}, 2417 (1999).

\bibitem{Vitali} D. Vitali and P. Tombesi, Phys. Rev. A, {\bf 59}, 4178 (1999). 

\bibitem{ELSE} C. A. Mead and D. G. Truhlar, J. Chem. Phys., {\bf 70}, 2284 (1984);
J. Moody, A. Shapere and F. Wilczek, Phys. Rev. Lett., {\bf 56}, 893 (1986);
H. Kuratsuji and S. Iida, Phys. Rev. Lett., {\bf 56}, 1003 (1986);
G. Delacr\'etaz et al, Phys. Rev. Lett., {\bf 56}, 2598 (1986).

\end{references}
\end{document}